\documentclass[aps,twocolumn,showpacs]{revtex4-1}
\usepackage[english]{babel}
\usepackage{graphicx}

\begin{document}
\title{Random packing of regular polygons and star polygons on a flat two-dimensional surface}
\author{Micha\l{} Cie\'sla}
 \email{michal.ciesla@uj.edu.pl}
 \affiliation{Marian Smoluchowski Institute of Physics, Jagiellonian University, 30-059 Krak\'ow, Reymonta 4, Poland}
\author{Jakub Barbasz}
 \email{ncbarbas@cyf-kr.edu.pl}
 \affiliation{Institute of Catalysis and Surface Chemistry, Polish Academy of Sciences, 30-239 Krak\'ow, Niezapominajek 8, Poland}


\begin{abstract}
Random packing of unoriented regular polygons and star polygons on a two-dimensional flat, continuous surface is studied numerically using random sequential adsorption algorithm. Obtained results are analyzed to determine saturated random packing ratio as well as its density autocorrelation function. Additionally, the kinetics of packing growth and available surface function are measured. In general, stars give lower packing ratios than polygons, but, when the number of vertexes is large enough, both shapes approach disks and, therefore, properties of their packing reproduce already known results for disks.
\end{abstract}
\maketitle
\section{Introduction}
Random packing of different objects on a two-dimensional surface became a very active field of study over three decades ago when it occurred that it can model irreversible adsorption processes, which are of great importance in life-sciences as well as in industry \cite{bib:Torquato2010,bib:Dabrowski2001}. The main focus of scientific investigations in this area was random packing of quite simple convex shapes like disks \cite{bib:Feder1980}, squares \cite{bib:Viot1990}, rectangles \cite{bib:Vigil1989}, and ellipsoids \cite{bib:Viot1992}. As experimental techniques have improved and the density of an adsorption layer can be measured more accurately, it occurs that such simple convex models of adsorbed particles can be insufficient to explain experimental results. Therefore, growing interest in packing of more complex, not only convex shapes has been recently observed \cite{bib:Adamczyk2012,bib:Jiao2008}. For example, it has been shown that a non-convex model of fibrinogen molecule can explain properties of adsorbed monolayers much better than a model based on spherocylinder shape \cite{bib:Adamczyk2010}. Recent studies suggest that particle shape, symmetry, and anisotropy are essential not only for obtained surface densities but also for kinetics of the adsorption process \cite{bib:Ciesla2014ring}. Even an imperceptibly small change of concave particle anisotropy can cause significant changes \cite{bib:Ciesla2014dim}. 
\par
Therefore, the aim of this study is to check how shape of an adsorbed particle, especially if it is convex or concave, influences main properties of their random packing. To provide a systematic approach to this problem, random packing of regular polygons and stars is analyzed. Both the shapes can have the same number of symmetry axes and, therefore, the only cause of any potential difference should be connected with either convex or concave shape of the particle. Moreover, with the increasing number of vertexes, both shapes are becoming closer and closer approximation of disks - the most extensively case studied so far. 
\section{Model}
\begin{figure}[htb]
\centerline{%
\includegraphics[width=0.8\columnwidth]{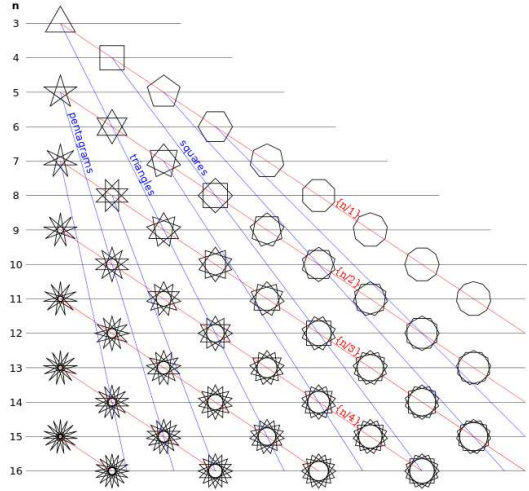}}
\caption{(Color online) Regular polygons and star polygons described by different Schl\"afli symbols \cite{bib:Wiki1}.}
\label{fig:stars}
\end{figure}

Regular polygon and star polygon can be described by the Schl\"afli symbol $\{n/k\}$, where $n$ is a number of regularly spaced points on its circumference and $k$ tells us that every $k$-th point should be connected with a line \cite{bib:MW1}. Note that for regular polygons $k=1$. For example, pentagon is described by $\{5/1\}$ symbol, whereas pentagram by $\{5/2\}$ (see Fig.\ref{fig:stars}).
In this study parameters $n$ and $k$ were restricted to $n\le 50$ and $k \in \{1,2,3,4,5\}$
\par 
Such shapes were thrown randomly on a flat and homogeneous surface according to random sequential adsorption (RSA) algorithm \cite{bib:Feder1980}, which iteratively repeats the following steps:
\begin{description}
\item[] a virtual particle is created on a surface. Its position and orientation were drawn from uniform probability distributions;
\item[] if the virtual particle does not overlap any of its closest neighbors it stays on a surface and holds its position and orientation until the end of simulation;
\item[] if there is an overlap the virtual particle is removed and abandoned.
\end{description}
The number of iterations $l$ is typically measured in dimensionless time units:
\begin{equation}
t_0 = l \frac{S_p}{S},
\end{equation}
where $S$ is the surface area and $S_p$ is the area covered by a single particle (polygon or star). Assuming that the distance between neighboring points in a regular polygon is a unit length, the $S_p$ for a regular polygon and a star polygon is given by:
\begin{equation}
S_p = \frac{n}{4}\left[\cot \left( \frac{\pi}{n} \right) - \tan\left( \frac{(k-1)\pi}{n} \right) \right].
\end{equation}
Here, the surface was a square of the side size from $250$ for the smallest particles ($n<10$) up to $1000$ for the largest ones ($n>20$).
The simulation was stopped when dimensionless time exceeded $t=10^5 t_0$. For each particle shape $\{n/k\}$, $10$ to $100$ independent simulations were performed to improve statistics. During simulations, the fraction of covered surface, known also as a coverage (or packing) ratio, was measured:
\begin{equation}
\theta(t) = N(t)\frac{S_p}{S},
\end{equation}
where $N(t)$ is the number of particles on a surface after the number of iterations corresponding to dimensionless time $t$.
\section{Results}
\begin{figure}[htb]
\centerline{%
\includegraphics[width=0.4\columnwidth]{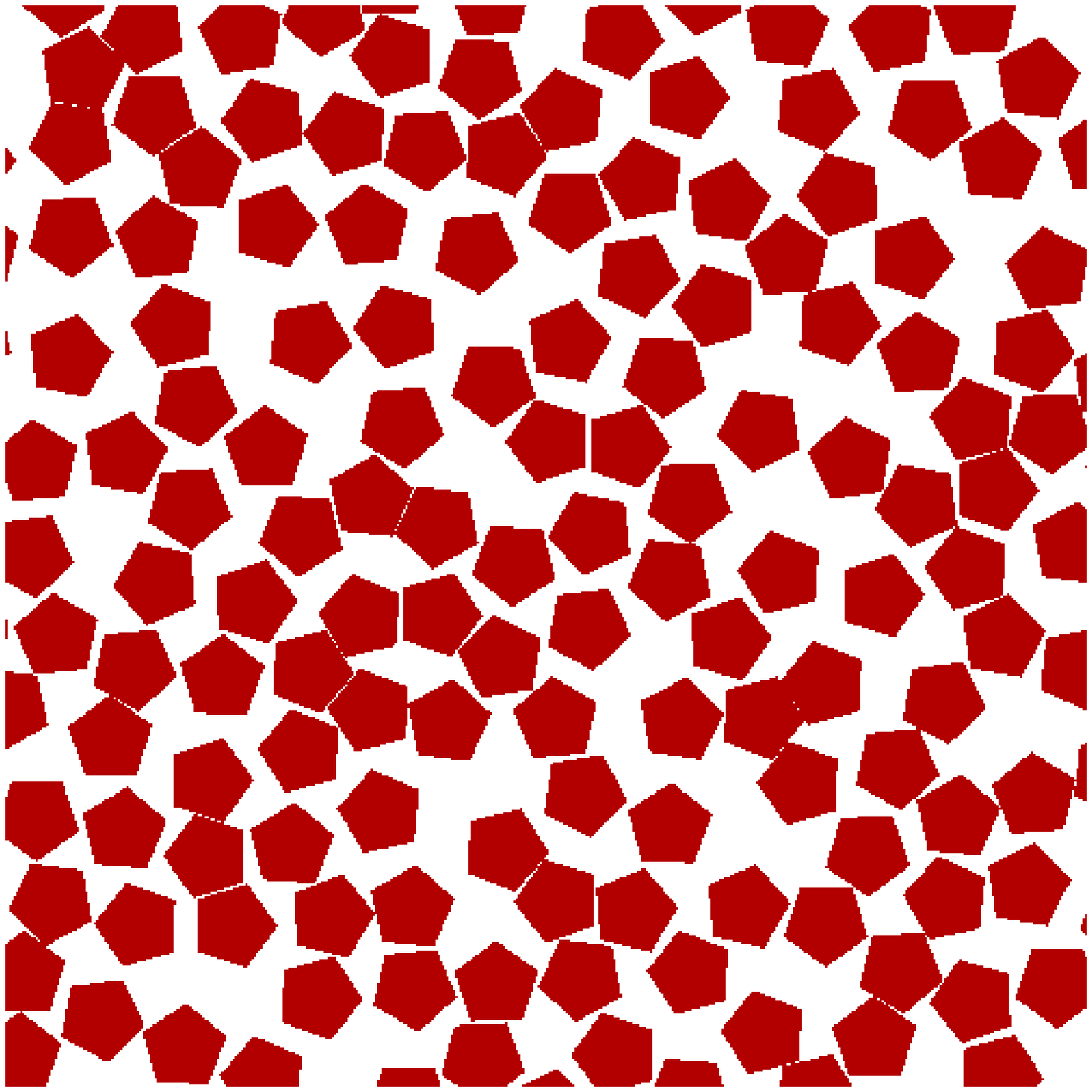}
(a)
\hspace{0.5cm}
\includegraphics[width=0.4\columnwidth]{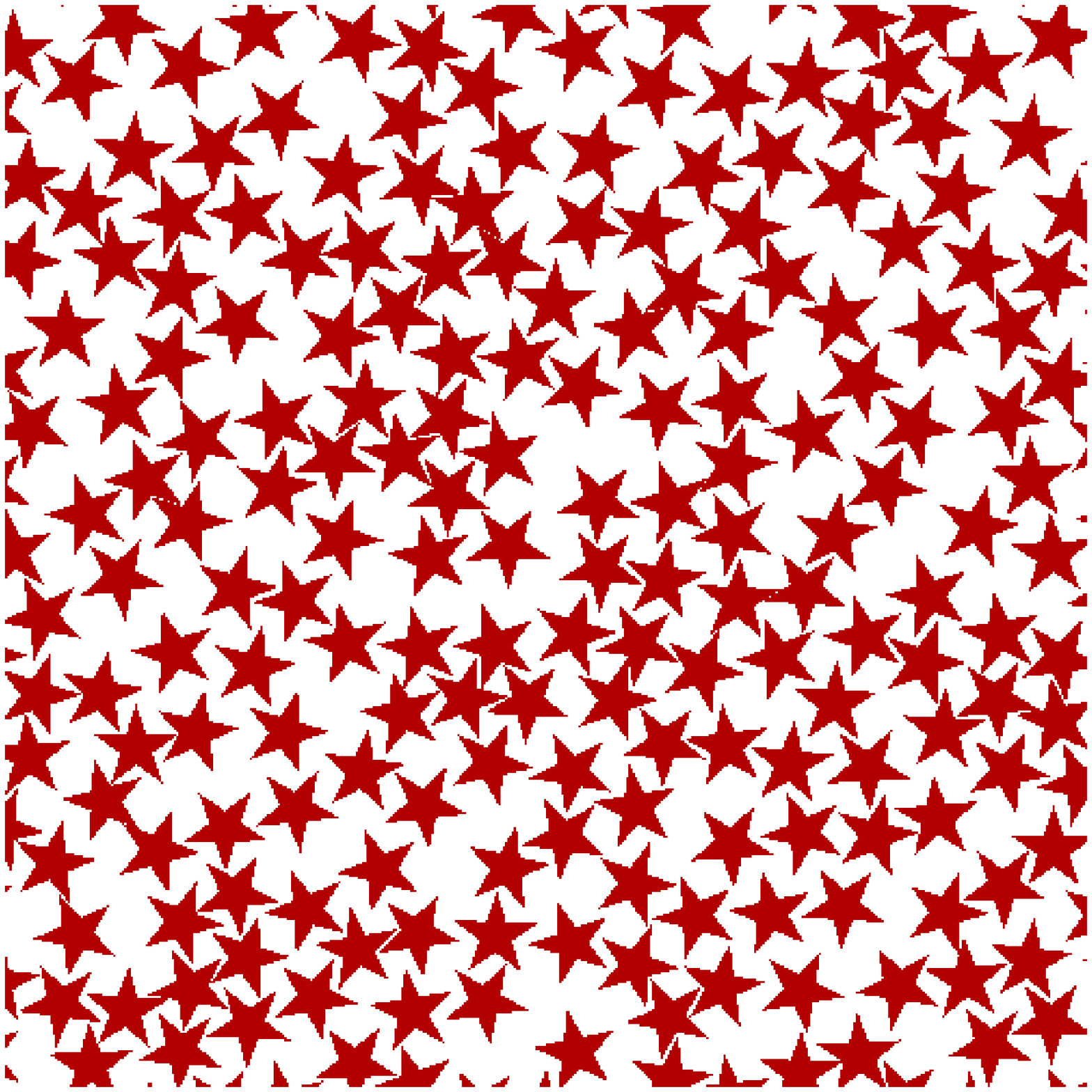}
(b)
}
\vspace{0.5cm}
\centerline{%
\includegraphics[width=0.4\columnwidth]{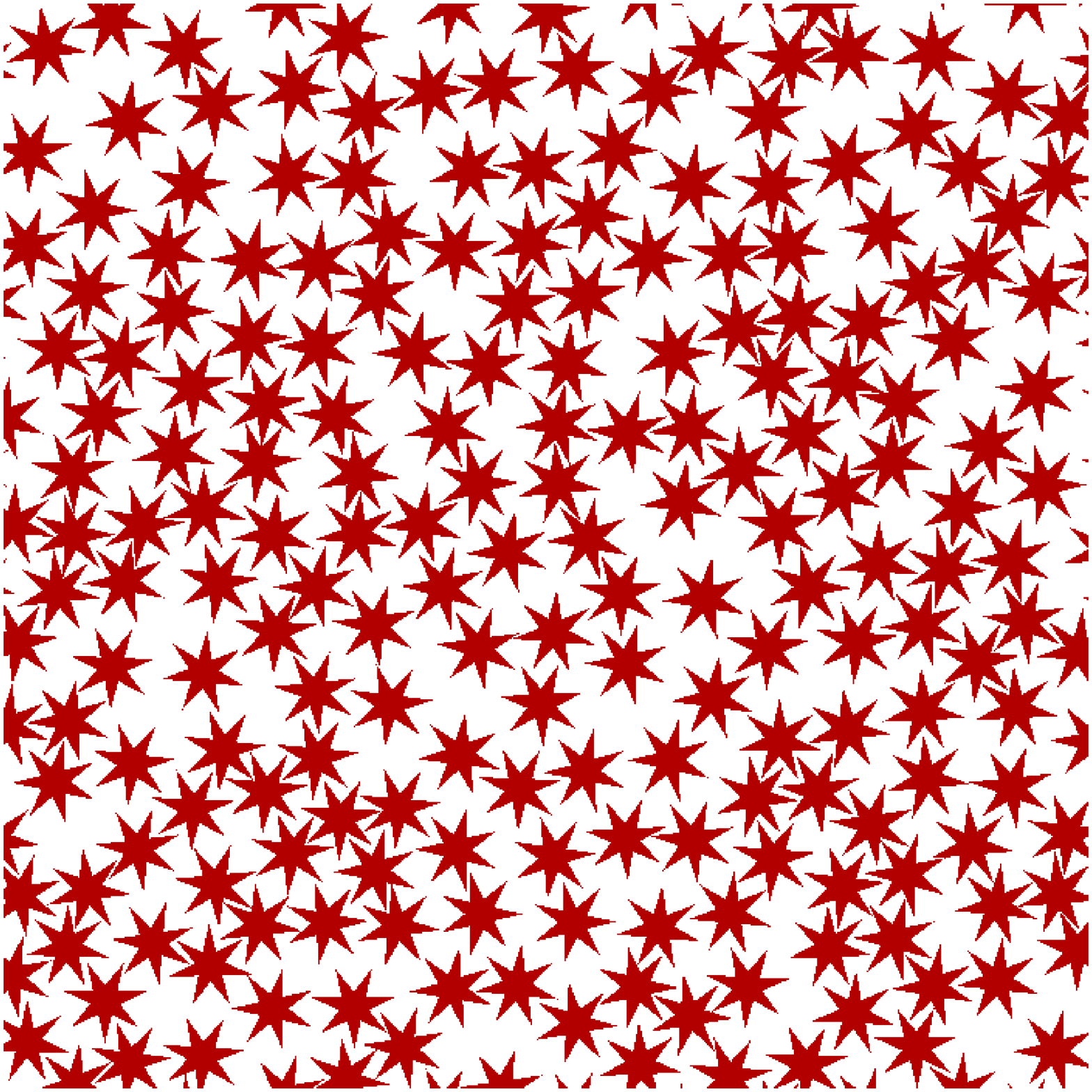}
(c)
\hspace{0.5cm}
\includegraphics[width=0.4\columnwidth]{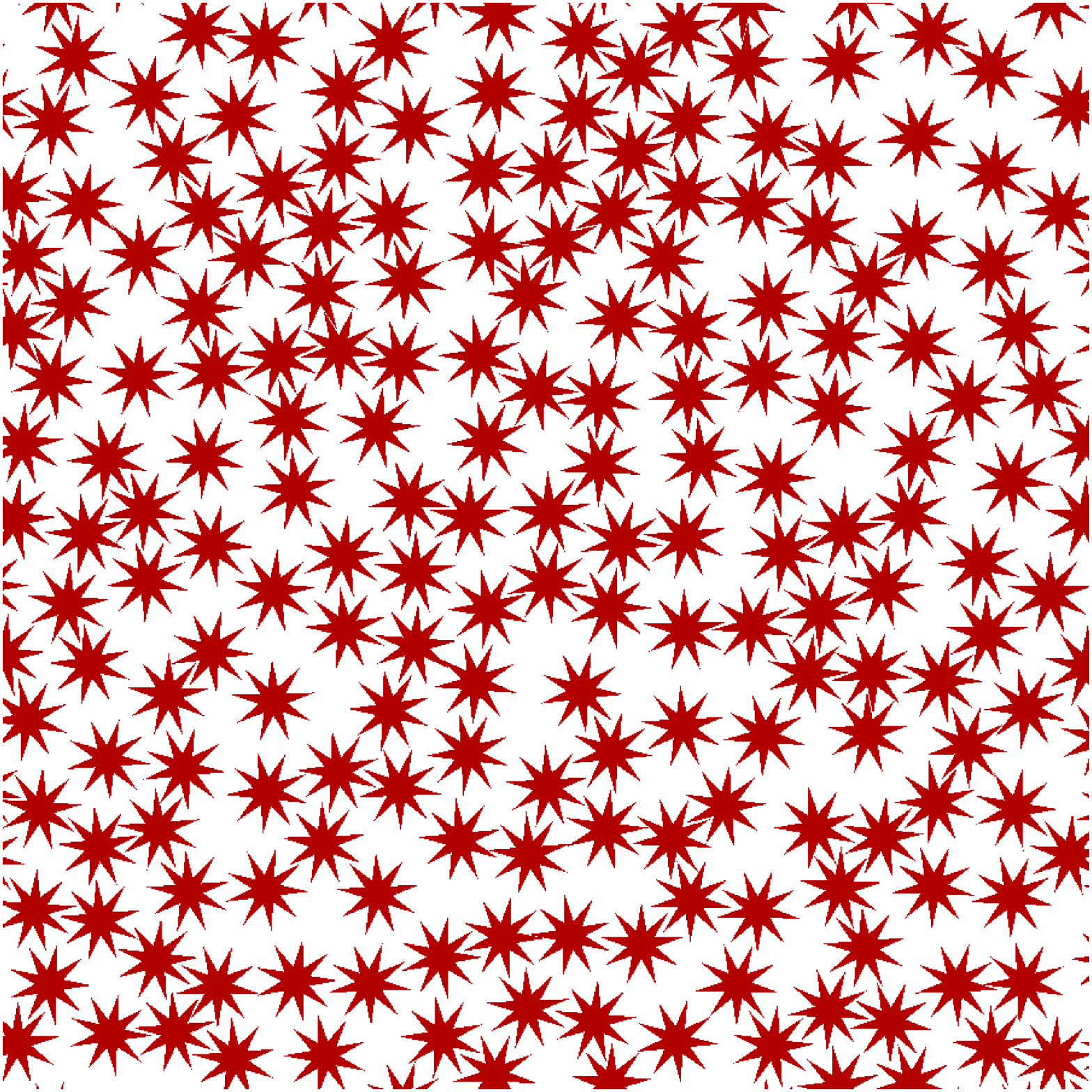}
(d)
}
\caption{Fragments of example random packings built of polygons $\{5/1\}$ (a), pentagrams $\{5/2\}$ (b), heptagrams $\{7/3\}$ (c), and enneagrams $\{9/4\}$ (d) at the end of simulation ($t=10^5$). They are rarer than saturated packings by approximately $0.5\%$.}
\label{fig:examples}
\end{figure}

Fragments of example packings are presented in Fig.\ref{fig:examples}.
As mentioned before, simulations were stopped after $10^5 t_0$ algorithms iterations. Therefore, packings at the end of numerical experiment, although very close, are slightly lower than saturated ones. To find saturated packings density the kinetics of RSA has to be analyzed.
\subsection{RSA kinetics}
It has been observed by Feder \cite{bib:Feder1980} and confirmed later by a number of numerical and analitical works \cite{bib:Pomeau1980,bib:Swendsen1981,bib:Privman1991,bib:Hinrichsen1986,bib:Viot1992} that the fraction of covered surface area scales asymptotically (for large $t$) with number of algorithm steps according to a power law
\begin{equation}
\theta(t) = \theta_{max} - A \, t^{-1/d}
\label{fl}
\end{equation}
where $\theta_{max}$ is the coverage ratio of saturated packing and $A$ is a positive constant. The value of parameter $d$ depends on both shape of packed objects and surface properties. For example, for RSA of disks on a two-dimensional plane, $d=2$; however, for RSA of rectangles, ellipsoids and other stiff but significantly anisotropic shapes on the same plane, $d=3$ \cite{bib:Viot1992,bib:Ciesla2013pol}. In general, it seems that parameter $d$ corresponds to a number of packed objects' degrees of freedom, which has been confirmed for random packing of hypherspheres in higher dimensions \cite{bib:Torquato2006,bib:Zhang2013}, not only the integral ones \cite{bib:Ciesla2012frac,bib:Ciesla2013frac}. 
\par
\begin{figure}[htb]
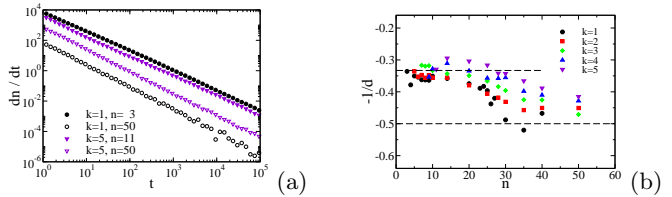

\centerline{%
\includegraphics[width=0.4\columnwidth]{kinetics1}
(a)
\hspace{0.5cm}
\includegraphics[width=0.4\columnwidth]{kinetics2}
(b)}
\caption{(Color online) Dependence of the increment of the number of packed objects on RSA dimensionless time for different $n$ and $k$ indexes (a). Dependence of the exponent in Eq.(\ref{fl}) on index $n$ of a regular polygon or a star polygon for different $k$ index (b). Horizontal dashed lines correspond to limits known for anisotropic objects ($d=3$) and for spherical ones ($d=2$).}
\label{fig:kinetics}
\end{figure}

For RSA of regular polygons and star polygons studied here, the power law (\ref{fl}) is fulfilled; however, the exponent $-1/d$ significantly depends on a particle shape (see Fig.\ref{fig:kinetics}).
For small number of vertexes $n$, parameter $d$ is approximately equal to $3$, which is the value known for anisotropic molecules. This observation is in agreement with results obtained for RSA of beads circle studied in \cite{bib:Ciesla2014ring}. When number of vertexes increases, the shape of particle approximates to disk (see Fig.\ref{fig:stars}) and, as expected, the parameter $d$ tends to $2$; however, there is no rapid transition between those two limits as it was observed in the case of beads circles \cite{bib:Ciesla2014ring} or generalized dimers \cite{bib:Ciesla2014dim}. The dependence of $d$ on $n$ is qualitatively the same regardless of $k$ index describing the type of star polygon, but it seems that larger $k$ corresponds to slightly higher $d$. It is in a good agreement with intuitive anticipation as the regular star $\{9/4\}$ is definitely more anisotropic than $\{9/2\}$ regular star or $\{9/1\}$ regular polygon.
\subsection{Saturated random packing}
\begin{figure}[htb]
\vspace{1cm}
\centerline{%
\includegraphics[width=0.8\columnwidth]{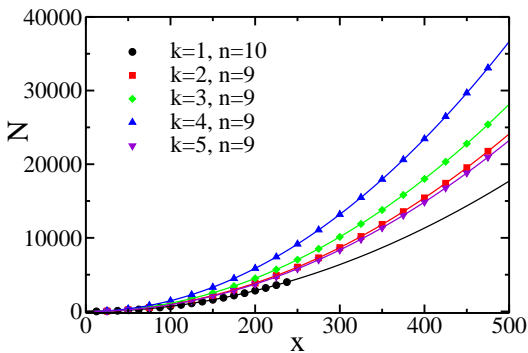}}
\caption{(Color online) Dependence of particle number in saturated random packing $N_{t\to\infty}$ on the square surface side size $x=\sqrt{S}$ for exemplary shapes. Dots represent data obtained from numerical simulations. Lines are quadratic fits: 
$N(k=1, n=10) = 0.23376 - 0.014518  x + 0.070824 x^2$;
$N(k=2,n=9) = 0.010649 - 0.0075317  x + 0.096439 x^2$;
$N(k=3,n=9) = -0.35072 + 0.018628  x + 0.11249 x^2$;
$N(k=4,n=9) = -2.9365 - 0.023171 x + 0.14663 x^2$;
$N(k=5,n=9) = 4.2778 - 0.051665 x + 0.092982 x^2$.
}
\label{fig:N-size}
\end{figure}

\begin{figure}[htb]
\vspace{1cm}
\centerline{%
\includegraphics[width=0.8\columnwidth]{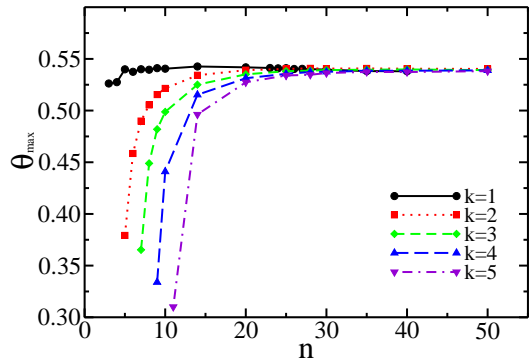}}
\caption{(Color online) Dependence of saturated random coverage ratio $\theta_{max}$ on the number of vertexes of a regular polygon or a regular star polygon for different index $k$. Dots represent data obtained from numerical simulations. Lines are shown to guide the eye.}
\label{fig:coverage}
\end{figure}

Having determined parameter $d$, relation (\ref{fl}) can be rewritten as $\theta(y) = \theta_{max} - A\,y$, where $y=t^{-1/d}$. Then, the density of saturated packing for finite collector size $\theta_{max}$ can be read from linear fit to $\theta(y)$ function in the point $y=0$. To find density of saturated packing for infinitely large surface the same procedure should be done for several finite, but different sized systems. As it is shown in Fig.\ref{fig:N-size} the dependence between approximated number of packed particles after infinite simulation time and square surface side length $x=\sqrt{S}$ is quadratic: $N_{t\to\infty}(x) = a x^2 + b x + c$. 
The existence of linear and constant term is connected with boundaries and finite size effects respectively. As the particles density is equal to $N/S$, for infinite system ($x\to\infty$) it is given directly by coefficient $a$. Therefore $\theta_{max}(t\to \infty, S\to \infty) = a S_p$. In the case of presented results these approximations make saturated densities not more than $1\%$ higher than one measured at the end of simulation. The dependence of $\theta_{max}$ on a number of vertexes of packed object is shown in Fig.\ref{fig:coverage}.
As expected, for large $n$, the saturated random coverage ratio $\theta_{max}$ approaches the value of $0.547$ known as saturated random coverage of spheres. For small number of vertexes and resulting higher shape anisotropy, coverage ratios are significantly smaller. Moreover, the dependence seems to be monotonic -- there is no clear maximum for moderate anisotropies like in the case of RSA of rectangles \cite{bib:Vigil1989}, ellipsoids \cite{bib:Viot1992}, and generalized dimers \cite{bib:Ciesla2014dim}.
The larger index $k$ is the lower saturated coverage is observed. It suggests that RSA of star polygons does not promote gears-like configuration when packing is saturated.
\subsection{Density autocorrelation function}
\begin{figure}[htb]
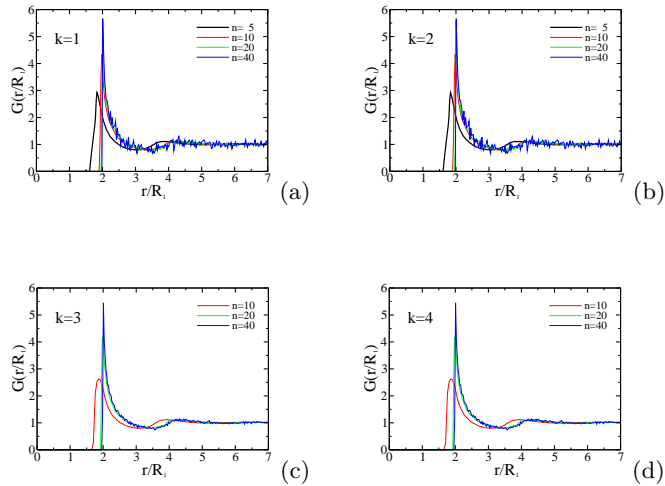

\centerline{%
\includegraphics[width=0.4\columnwidth]{cor1}
(a)
\hspace{0.5 cm}
\includegraphics[width=0.4\columnwidth]{cor2}
(b)
}
\vspace{1.0cm}
\centerline{%
\includegraphics[width=0.4\columnwidth]{cor3}
(c)
\hspace{0.5 cm}
\includegraphics[width=0.4\columnwidth]{cor4}
(d)
}
\caption{(Color online) Normalized density autocorrelation function for regular polygons and regular star polygons. Different graphs correspond to different type of star: $k=1$ (a), $k=2$ (b), $k=3$ (c), and $k=4$ (d). Different lines correspond to different number of polygon or star polygon vertexes: $n \in \{10, 20, 40\}$, and additionally $n=5$ for $k=1$ and $k=2$.}
\label{fig:cor}
\end{figure}

\begin{figure}[htb]
\centerline{%
\includegraphics[width=0.6\columnwidth]{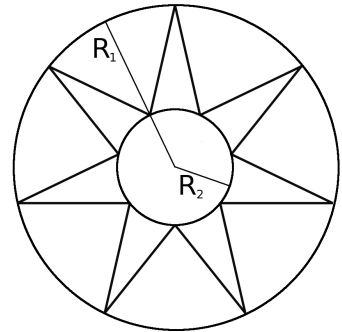}
}
\caption{Circumcircle and incircle radiuses of an example regular star polygon.}
\label{fig:r1r2}
\end{figure}

Saturated random coverage $\theta_{max}$ contains information about mean density of packed particles. To obtain additional information about their structure, density autocorrelation function can be studied. It is defined as a normalized probability density $p(r)$ of finding two different object in a given distance $r$:
\begin{equation}
G(r) = \frac{p(r)}{2\pi r \rho},
\end{equation}
where $\rho$ is the mean density of packed objects. Note that such normalization leads to $G(r\to \infty) = 1$ To compare density autocorrelation function for different shapes, it should be normalized to take into account different sizes of polygons and stars. Therefore, instead of $G(r)$, $G(r/R_1)$ is discussed, where 
\begin{equation}
R_1 = \frac{1}{2 \sin\left(\frac{\pi}{n} \right)}
\end{equation}
Results are shown in Fig.\ref{fig:cor}
As previously, for large number of vertexes ($n=40$), regardless the value of $k$, the density autocorrelation function is almost identical to the one for disks: it has logarithmic singularity at $r/R_1 = 2$ \cite{bib:Swendsen1981} and decays superexponentially for larger distances \cite{bib:Bonnier1994}. For relatively small $n$, the first maximum of $G(r/R_1)$ is significantly below $r/R_1 = 2$, especially for star polygons. To explain this, it is worth to mention that there are two characteristic distances describing a regular polygon or a star polygon. First is the $R_1$, which is a radius of circumscribed circle of a polygon. The second is a radius of inscribed circle of a regular polygon or a star polygon (see Fig.\ref{fig:r1r2}):
\begin{equation}
R_2 = \frac{1}{2} \left[\cot \left( \frac{\pi}{n} \right) - \tan \left( \frac{(k-1)\pi}{n} \right)\right].
\end{equation}
The minimal distance between objects is $R_1+R_2$, which is, especially for small $n$, significantly lower than $2 R_2$. Of course for large $n$, $R_1 \to R_2$ and, therefore, $R_1 + R_2 \to 2R_1$, which is equal to the lowest possible distance between two disks of radius $R_1$.
\subsection{Low coverage limit and viral coefficients}
\begin{figure}[htb]
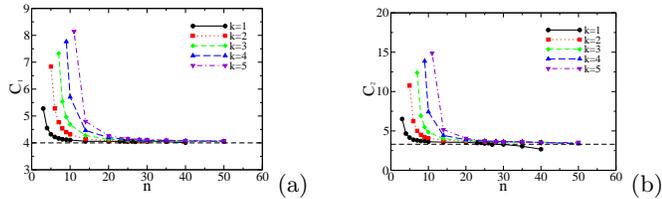

\vspace{1.0cm}
\centerline{%
\includegraphics[width=0.4\columnwidth]{c1}
(a)
\hspace{0.5cm}
\includegraphics[width=0.4\columnwidth]{c2}
(b)
}
\caption{(Color online) $C_1$ (a) and $C_2$ (b) coefficients dependence on the number of polygon vertexes $n$. Different lines correspond to different polygon or star polygon types: $k \in \{1,2,3,4,5\}$. Dots are the data obtained from numerical simulations. Dashed line corresponds to value known for disks: $C_1 = 4$ and $C_2 = \frac{6\sqrt{3}}{\pi}$. Solid lines are drawn to guide the eye.}
\label{fig:c1c2}
\end{figure}

The probability of successful addition of a subsequent particle to a packing under RSA protocol is described by Available Surface Function (ASF). For low packing ratio, it can be expanded in a Taylor series
\begin{equation}
ASF(\theta) = 1 - C_1 \theta + C_2 \theta^2 + o(\theta^2).
\end{equation}
Parameter $C_1$ corresponds to the area blocked by a single particle, whereas $C_2$ is the mean cross-section of those areas for two neighboring particles. Parameters $C_1$ and $C_2$ are directly related to coefficients in viral expansion for equilibrium state of a packing
\begin{equation}
\frac{p}{k_B T} = \rho + \rho B_2(T) + \rho^2 B_3(T) + o(\rho^2),
\end{equation}
where $p$ is pressure, $T$ is temperature, $\rho$ is particle density, and $k_B$ is the Boltzmann constant. For example, it can be shown that $C_1 = 2B_2$ and $C_2 = 2B_2^2 - \frac{3}{2}B_3$ \cite{bib:Ricci1992,bib:Schaaf1989}. It is worth to note, that for disks $C_1 = 4$ and $C_2 = \frac{6\sqrt{3}}{\pi}$. Moreover, analitical formula for $C_1$ for regular polygons was found by Pozhar et al. \cite{bib:Pozhar2003}. $ASF(\theta)$ function can be directly measured during RSA simulations. It can also be measured in adsorption experiments as it is related to density fluctuations of adsorbed particles \cite{bib:Schaaf1995,bib:Adamczyk1996}. Here, $C_1$ and $C_2$ coefficients were measured by fitting the quadratic function to $ASF(\theta)$ for $\theta / \theta_{max} < 0.2$. Results are shown in Fig.\ref{fig:c1c2}.
Presented plots correspond well to saturated random packings shown in Fig.\ref{fig:coverage}. The relatively high blocked surface ($C_1$ parameter) for small $n$ results in low saturated packing ratio. For large $n$, all plots approach $C_1=4$ known for disks; also, packing ratio tends to value known for disks. The same could be stated about $C_2$'s dependence on the number of polygon vertexes.
\section{Summary}
The saturated random packing ratios of concave regular star polygons are lower than those of convex regular polygons because concave particles block almost the same area as regular polygons while their surface is smaller. This effect is stronger for small number of particle vertexes as well as for larger index $k$ describing the type of a star. For large number of polygon vertexes, the shape of regular polygons and regular star polygons approximates to disk thus the main properties of random packings such as saturated ratio, RSA kinetics, density autocorrelation or thermodynamic characteristics are almost the same as those for disk random packings.
\section*{Acknowledgments}
This work was supported by Polish National Science Center Grant No. UMO-2012/07/B/ST4/00559.

\end{document}